# New Visual Cryptography Algorithm For Colored Image

Sozan Abdulla

**Abstract:** Visual Cryptography is a special encryption technique to hide information in images, which divide secret image into multiple layers. Each layer holds some information. The receiver aligns the layers and the secret information is revealed by human vision without any complex computation. The proposed algorithm is for color image, that presents a system which takes four pictures as an input and generates three images which correspond to three of the four input pictures. The decoding requires only selecting some subset of these 3 images, making transparencies of them, and stacking them on top of each other, so the forth picture is reconstructed by printing the three output images onto transparencies and stacking them together. The reconstructed image achieved in same size with original secret image.

**Index term:** visual cryptography, secret sharing, primitive color.

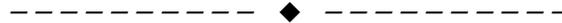

## 1 INTRODUCTION

Visual cryptography is a popular solution for image encryption. Using secret sharing concepts, the encryption procedure encrypts a secret image into the shares(printed on transparencies) which are noise-like secure images which can be transmitted or distributed over an untrusted communication channel. Using the properties of the HVS to force the recognition of a secret message from overlapping shares, the secret image is decrypted without additional computations and any knowledge of cryptography [1].

Visual cryptography is proposed in 1994 by Naor and Shamir who introduced a simple but perfectly secure way that allows secret sharing without any cryptographic computation, which they termed as Visual Cryptography Scheme (VCS). The simplest Visual Cryptography Scheme is given by the idea of A secret image consists of a collection of black and white pixels where each pixel is treated independently[2].

There are many algorithm to encrypt the image in another image, but a few of them have been in visual cryptography for color image. In this paper, the different approach have been produced for the visual cryptography for color image, the proposed algorithm splits a secret image into three shares based on three primitive color components.

Section 2 in this paper give a review of the literature and research in this field and it concerns with the main fundamental concept to understand the idea of visual cryptography of (black and white) and color image and theoretical foundation for each algorithm. In section 3, we present the problem under consideration and introduce the proposed algorithm and constructing of new algorithm is explained.

Section 4, the experimental results are discussed in detail.. Finally, Section 5 and 6 summarizes the conclusions and future work.

## 2 VISUALCRYPTOGRAPHY MODEL

A printed page of ciphertext and a printed transparency (which serve as a secret key). The original clear text is revealed by placing the transparency with the key over the page with the cipher, even though each one of them is indistinguishable from random noise. The model for visual secret sharing is as follows. There is a secret picture to be shared among n participants. The picture is divided into n transparencies (shares) such that if any m transparencies are placed together, the picture becomes visible. If fewer than m transparencies are placed together, nothing can be seen. Such a scheme is constructed by viewing the secret picture as a set of black and white pixels and handling each pixel separately[1].

There are some researches that deal with visual cryptography for color images. In this section, we will explore several representative work over the years. [3] discussed the visual cryptography scheme which reconstructs a message with two colors, by arranging the colored or transparent subpixels, they only discussed construction of (k,n)-threshold scheme where a subset X 2 $\in$ $\wp$ is a qualified set if and only if |X| = k. Koga et al. devised a lattice-based (k, n) scheme[4]. The approach by [5] is basically similar to Koga's. Both approaches assign a color to a subpixel at a certain position, which means that displaying m colors uses m−1 subpixels. The resulting pixels contain one colored subpixel and the rest of the subpixels are black. Therefore the more colors are used, the worse the contrast of the images becomes significantly. Their approaches cannot be applied to the extended visual cryptography, either. Rijmen and Preneel talked about enabling multicolors with relatively less subpixels(24 colors with m = 4)[6]. However each sheets must contain color random images, which means applying this approach to the extended visual cryptography is impossible.[ 7]

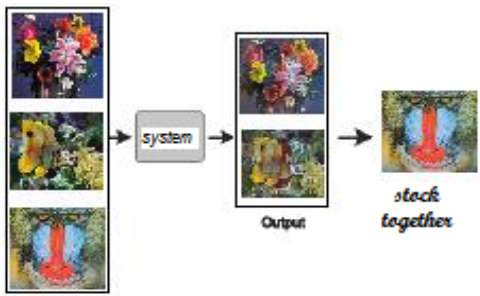

Figure 1:  visual cryptography system

In visual cryptography system the pixels  of the image to be encrypted  can be applied to the image in  different  manner.  There  is  a  set  of  n participants(image),   and  the  secret  image   is divided and encoded into n shadow images called shares Each participant encrypted by  one share, k out of n participants are needed to combine shares and see secret image, some time k-1  of shares can not reveals  information about secret image.

The technology makes use of the human vision system to perform the OR logical operation on the superimposed pixels of the shares. When the pixels are small enough and packed in high density, the human vision system will average out the colors of surrounding pixels and produce a smoothed mental image in a human's mind. For example, a block of 2 × 2 pixels will be viewed as a gray-like dot as the two black pixels and the two nearby white pixels are averaged out. If we print the 2×2 pixel blocks shown in Fig. 2 separately onto two transparencies and superimpose them. This effect is equivalent to performing a pixel-wise OR logical operation on each of the four pairs of pixels between these two transparencies. The result is shown in Fig. 3. One of the unique and desirable properties of VCS is that the secret recovery process can easily be carried out by superimposing a number of shares (i.e. transparencies) without requiring any computation[8].

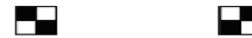

Figure 2 : Two 2 × 2 pixel blocks

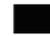

Figure 3 : Superimposed Image

# 3 EXPERIMENTAL DESIGNS

Every single pixel in secret image is splited into subpixels , that can still perceived them as one pixel by Human vision system. The security of each share depends crucially on the color composition of the original secret image. To recovering a secret image, it is required that the cover image should at least be able to determine the shape or pattern of the original secret image, which is able to determine the boundary between two distinct color regions in the image. In this paper, for the problem of interest here, assuming  an input 24-bit bitmap color image which each 3-byte sequence in the bitmap array represents the  relative intensities of red, green, and blue, respectively  for image  sized 256×256 RGB pixel for hiding  Baboon  image.

In the following, the steps of the proposed algorithm and illustrate it using an example where the original secret image is a 24-bit color baboon image shown in Fig.4. The encryption process consists of determining the arrangements of transparent subpixels on each sheet according to the pixel transparencies,

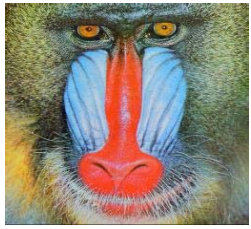

Figure 4: Original secret image

**Step** 1. The proposed algorithm first decomposes the original image into three primitive-color images under the subtractive model, namely, C (Cyan), M (Magenta) and Y (Yellow). Fig. 5 shows the three primitive color components of the baboon image, where each image has 256 levels of the corresponding primitive color, and each pixel represented by three bytes. Converting to (C,M, Y ), where C,M, Y $\in$ {0- 255}.

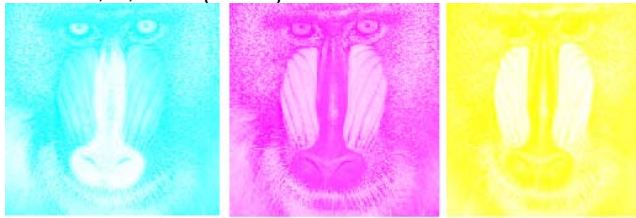

Figure 5: Primitive Color (C, M, Y ) Components

**Step** 2 For each pixel of the C primitive $P_{i,j}$ will be store it in the other image for reducing the $P_{i,j}$ value by holding ¼ $p_{i,j}$ Fig.6 show the color of a pixel in the original baboon secret image in terms of primitive cyan, magenta and yellow with reducing value, respectively.

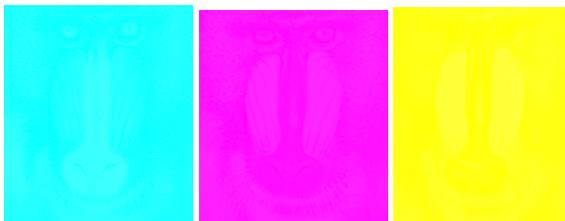

Figure 6: Reducing pixels value of (C,M,Y)

**Step** 3 : Selecting the cove image to encrypt on it the secret image depends on minimum difference between the value of pixel, there are three primitive color, therefore we need three cover image to generate three share .

**Step** 4: In this step, three shares, namely a, b and c shares are generated each share contains apart of secret image. To generate these shares, the C, M and Y the reduction primitive-color images of the original secret image are read in pixel by pixel and mixing(OR operation ) together with ¾ pixel of cover image shown in Fig .7. For example if the pixel value of primitive color equal 00000011 and the value of the cover image is 01101100 the out put pixel will be 01101111 by OR operation. So, each share (a, b, and c) hold part a primitive color for secret image, share (a) which is baby image holds C primitive color , share (b) flower image holds M primitive part, and share(c) fish image hold Y primitive color , selecting of the cover image depending on minimum differences between pixel in cover image and primitive color in secret image.

**Step** 5: Repeat step 1 to 4 for all pixel from original image

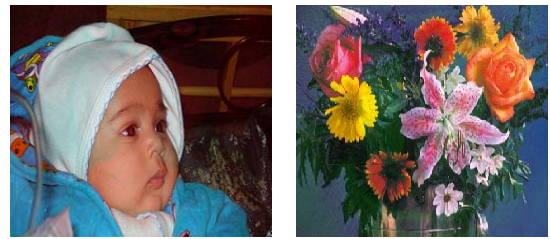

(a)                          (b)

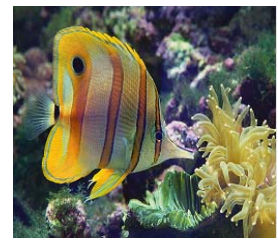

(c)

Figure7 : The Generation of C, M, Y Shares

To share a secret pixel with color (C,M, Y) in cover image depending on the value of threshold for selection the proper image. It is found that the shares ( a,b,c) reveal the information about the secret image . To recover the secret original baboon image it is necessary to removing the unwanted color in each share by the operation 255 minus ¾ CPi,j, where CPi,j $\in$ image (a,b,c).

## 4  EXPERIMENTAL RESULT

In this paper, the number of pixel in the decoded image  is same as in the original secret image because the pixel p  not expanded. The size of secrete image must be smaller or equal to the cover images. After testing many different images from different colors and  in resolution it was observed that the proposed algorithm could not take dark image significantly with high contrast and which then generate the unclear share, and they appears as images corrupted with large amount of noisy and it will be easy to detect.

We have added  some processing elements after recovering the secret image to change the quality of color to be more clear and give the similar original view. This is to be done  after removing unwanted color from the share and before stacker the shares together.  The way of  this processing is changing pixel value to Pi,j  multiplying by 4 to get full  value from 255 , so for secret image  will reveal the secret completely in shares in case of with the previous processing.

The way of reduction in original pixel and subtractions of the original pixel with previous shares pixel give more sensitive result and with better color  quality.  Our results suggest that the security of the image depends critically on the color composition of the  secret image  and distribution of the original secret image.

## 5  FUTURE WORK

The future work is to improve the contrast and produce more clear the resultant secret image. Further extend this work to use this technique with othe format of color images. Also consider 3D images for creating the shares that have partial secret and reveal that secret by stacking to each other. In reality, however, such ideal subtractive color mixture is unlikely due to the properties of ink, transparencies, etc. It needs to establish a sophisticated color mixing model for the extended visual cryptography with better color quality.

## 6  CONCLUSION

In this paper, provided a detailed security visual cryptography  new algorithm  for 24-bit bitmap color image . After splitting the image to the number of  the shares and  applying some recoloring method  according to the color mixture groups we identified  the adversary may find out some useful information such as the shape or pattern of the original secret image. The number of images are  tested  to hiding the  part secret image and it is  found that the image with color near to the primitive C (Cyan), M (Magenta) and Y (Yellow). In particular, it can find out the shape of the original secret image with   %25  in each share and the original     secret image is color image and the adversary has acquired three of the C, M and Y shares. The winning chance of the adversary decreasing so it get more security because to recover the original image it need to stacker all the three shares , and each share may only be able to retrieve partial information of the original secret image.

Our results suggest that the security of the scheme depends critically on the color composition and distribution of the original secret image.

### ACKNOWLEDGMENT

I would like to express my gratitude to the anonymous reviewers  for their comments to improve the quality of  this paper.

### REFERENCES

[1] Talal Mousa Alkharobi, Aleem Khalid  2003. *New Algorithm For Halftone Image Visual Cryptography*, Alvi King Fahd University of Pet. & Min. Dhahran.

[2[  JIM CAI 2003. *A Short Survey on Visual Cryptography Schemes,*www.wisdom.weizmann.ac.il/naor/ PUZZLES /visual.html.

[3] Nakajima, M. and Yamaguchi, Y.2002. *Extended Visual Cryptography for Natural Images*, WSCG02, 303.

[4]  M. Naor and A. Shamir,1996.  *Visual cryptography ii: Improving the contrast via the cover base*. Theory of Cryptography Library, (96-07).

[5] Hiroki Koga and Hirosuke Yamamoto,1998. *Proposal of a lattice-based visual secret sharing scheme for color and gray-scale images*. IEICE

Transaction on Fundamentals, E81-A(6):1262–1269.

[6] E.R.Verheul and H.C.A.van Tilborg,1997. *Constructions and properties of k out of n visual secret sharing schemes*. Design Codes and Cryptography, 11(2):179–196.

[7] V. Rijmen and B. Preneel,1996. *Efficient colour visual encryption or shared colors of benetton. presented at EUROCRYPT,* 96 Rump Session, available as http://www.iacr.org/conferences/ec96/rump/preneel.ps.

[8] Bert W. Leung, Felix Y. Ng, and Duncan S. Wong, 2007. *On the Security of a Visual Cryptography Scheme for Color Images,* (RGC Ref. No. CityU 122107).